\documentstyle[11pt]{article}

\begin{document}

\author{Nestor Caticha and Osame Kinouchi \\
Instituto de F\'\i sica, Universidade de S\~ao Paulo\\
CP66318, CEP\ 05315-970, S\~ao Paulo, SP\ Brazil}
\title{Time ordering in the evolution of information processing and modulation
systems}
\date{}
\maketitle

\begin{abstract}
The ideas of optimization of learning algorithms in Artificial Neural
Networks are reviewed emphasizing generic properties and the online 
implementations are interpreted from a biological perspective. 
A simple model of
the relevant subsidiary variables needed to improve learning in artificial feedforward networks and the `time ordering' of the appearance of the respective information processing systems is proposed. 
We discuss the possibility that these results might be relevant in other contexts,
not being restricted to the
simple models from which they stem. 
The analysis of a few examples, which range from the lowest cellular scale to the macroscopic level, suggests that similar ideas could be applied to biological systems. 
\end{abstract}

\section{Introduction}

\subsection{Evolution and Optimization}

In the study of learning processes in artificial systems, the search for
general results can be pursued by concentrating on the statistical mechanics
of simple models
(Watkin, Rau and Biehl 1993). 
Rather than being interested in their peculiarities, the
aim is to unearth properties that, by recurrently appearing in several of
those models, may represent candidates of that sought after generality. The
ubiquity of these features may be an indication of their importance in more
complex systems, not amenable to an analytical approach, and thus help in
suggesting what are the important macroscopic variables in these systems.

The definition of efficiency of an artificial neural network (ANN) depends
on the task for which it has been defined. While criteria such as rote
memorization, generalization ability or ease of training may be used to
label and judge an ANN; adaptiveness, biological plausibility 
or implementation
possibilities can also be relevant parameters. The fact that a given ANN
scores well in one or other particular area may be enough to permit its
survival as a useful object of study. The construction of new ANN's from
scratch or the evolution to different artificial machines from previous
ones, will not be seen as the teleological drive towards the perfect
machine, for perfectness is not defined in this multidimensional value
space. This evolution leads, instead of a single line or lineage,
rather to the construction of a multibranched system.

Not to be thought of as a one dimensional process, this tree or bush can
nevertheless be used to define a single lineage process. Starting from the
tip of one branch and going backwards through the tree, a trajectory can be
distinguished from the neighboring paths that may eventually converge with
it as the historical path is retraced. Suppose this backward trajectory is
painted in a different color from other separating branches. Now we have an
evolution path for which there is at least a pseudo teleological dynamics.
Some quality measure, call it ${\cal Q}$ can be defined along this path, and
as the path is traveled along the forward time direction, then a drive
towards better ${\cal Q}$ can be identified.

In a competitive environment, the capacity to appropriately deal with and
efficiently process information may contribute to give an individual the
necessary fitness to be successful. While the build up of better information
processing devices is not the generic objective of naturally evolving
organisms, it is not totally indefensible that a painted line in a biological
evolution tree can be identified where the quantity ${\cal Q}$ is, at least
loosely, associated with the capacity to deal with information in a certain
specific manner.  
One of the most fundamental information processing capabilities
is the extraction of statistical regularities from the 
environment, i.e. statistical inference, 
while other such as memory or sensory processing  may be thought important
inasmuch as they contribute to the former and enhance the organism's 
predictive capacity.

In the last few years several papers have addressed the problem of
determining optimal generalization learning algorithms in ANN
(Kinouchi and Caticha 1992, 1993, 1993b, 1996, Biehl and Schwarze 1993, Watkin 1993, Copelli and Caticha 1995, Biehl, Riegler and Stechert 1995, Copelli, Kinouchi and Caticha 1996, Simonetti and Caticha 1996, Van den Broeck and Reimann 1996, Opper 1996, Opper and Winther 1996, Copelli {\it et al} 1997, Winther, Lautrup and Zang 1997)
The idea of
optimization of learning algorithms, whether applied to on or offline 
learning, in a
supervised or not manner, is based on the fact that a given
machine will be expected to perform satisfactorily in a rather restricted
environment. Although these algorithms may turn out to be
somewhat adaptive \footnote{ Tolerance to small changes in the
distribution of examples, drift in the underlying rules it tries to infer,
or to changing levels of corruption of the data by noise.}, it is unreasonable
to expect the same algorithm to be optimal under general conditions.
However, the specification of a restricted set of environmental variables
defines the learning scenario and makes the optimization problem well posed.
The learning scenario will define the ``evolutionary pressures'' which will
mold the learning algorithms. ${\cal Q}$ is here identified with the generalization ability.

The object of this communication is to deal with the problem of
optimization, first by quickly reviewing results obtained in relation to ANN
learning from examples
and then by giving an interpretation, from a biological point of view, of
several features that appear to be characteristic of optimized learning
algorithms. This is done in the hope that, if the features reflect
properties of learning in general, rather than simply showing effects
restricted to the particular chosen scenarios and architectures, then a
simple model of the time evolution towards more sophisticated biological
information processing can be suggested. The main point we want to stress is
that a `time' ordering can be seen to arise in the complexification of the
learning algorithms due to the importance of behavioral variables in the
efficient modulation of synaptical plasticity.

\subsection{Learning in the presence of Partial Information.}

The optimal learner relies on several auxiliary quantities that describe 
the (joint) probability distribution of examples, that is 
the learning scenario. 
In the absence of information on one or more such quantities a
full optimization cannot be carried out. Nevertheless, in the presence of partial information (sub)optimal algorithms can be found. Will the inclusion of one more
such variable in the `information pool' always lead to an increase in  ${\cal Q}$? Not necessarily, for
it may depend on which other variables are available. For example, consider two
auxiliary quantities $A$ and $B$.
Let $A^{-}$ represent that $A$ is missing from the information pool . 
Appropriate choices, for $A$ and $B$ can be made such that  
${\cal Q}(A^{-},B^{-}) = {\cal Q}(A,B^{-})<{\cal Q}(A^{-},B)<{\cal Q}(A,B)$.  

 We now imagine the
steps towards constructing the necessary hardware by some sort  of evolution.
A new piece of hardware that measures $A$ will not be useful and therefore
not included if the
necessary hardware to measure $B$ is not yet present. It can be said that $B$ potentiates $A$ as useful information. Evolution can take a path $(A^{-},B^{-}) \rightarrow (A^{-},B)\rightarrow(A,B)$, but not $(A^{-},B^{-}) \rightarrow (A,B^{-})\rightarrow(A,B)$.

This is what we mean
by a `time ordering' in the appearance of the different moduli that measure
the several relevant variables. It is an argument for at least time ordering
in the built up of functional modularity, but might be useful also for physical modularity development in the presence of some functional localization. 

Whether a similar claim holds for the
sequential construction of information processing moduli in biological
systems evolving under natural selection, is a very interesting but still
not clear proposition. We want to present a simple example to show that
this, in fact, may be the case.

\section{Optimal Learning in a class of ANN}

\subsection{Modulated Hebbian-like Learning}
\label{spuro}

We review results for a class of feedforward ANN where optimization has been
previously studied in the case of supervised learning.
It includes the simple perceptron
with continuous or binary weights, boolean or linear output, boolean reverse
wedge perceptron, tree parity and tree committee machines. For optimal unsupervised learning see (Van den Broeck and Reimann 1996).

The functions ($f:{\cal R}^N\Rightarrow {\cal R}$ or $\{\pm 1\}),\ $these
machines implement, depend on a set of $N$ parameters or coupling weights $%
\{J_i\}$ which are inspired by synaptic efficiencies in a biological neural
system and are supposed to play a similar role. The object of learning is to
modify the set of weights of a student network in order to approximate a
function $f_0$, unknown except for the information contained in the learning
set ${\cal L=}\{{\bf S}^\mu ,\sigma^\mu_b \}_{\mu =1,...P}$, i.e. , the
values $\{\sigma^\mu_b \}$ of the function at $P$ instances of the inputs $\{%
{\bf S}^\mu \}.$ The outputs$\{\sigma^\mu_b \}$ could even be corrupted by
some noise process. Optimization will have, in this work, the aim of
maximizing the ability of generalization, i.e. rule inference. While optimal
off-line learning has also been studied,
we concentrate in what follows in the
properties of on-line learning, because of its more biological appeal. We
consider the case of single online presentation of examples with no iteration.
These conditions are introduced in order to have manageable models from an
analytical point of view, but are not responsible for the properties we want
to discuss, which are still present under more general conditions or in
other more complex models.

For simplicity we discuss the simplest of all feedforward networks, the
single layer boolean perceptron, nevertheless the results are representative
of what is found for the other architectures mentioned above. The output $ 
\sigma_h $ of the perceptron is given by the sign of the 
post-synaptic field\footnote{
All vectors are written in boldface, e.g. ${\bf J,}$ while their
lengths, e.g. $J$ are not.} $h=
{\bf S.J}/J$, and the function $f_0$ it is learning is an
ANN\ of similar architecture, a teacher perceptron with an unknown set of $N$ real
weights $\left\{ B_i\right\} $. We will call $b={\bf S.B}/B,$ the
teacher's field. For these machines the generalization error $e_g$,
which measures the probability of disagreement between teacher and student
perceptrons is a monotonically decreasing function of the overlap $\rho =%
{\bf J.B}/JB$, for the perceptron: $e_g=\frac 1\pi \arccos \rho $.

During learning, the presentation of a new example to a network induces a
change $\Delta J_i$ in each `synaptic' weight $J_i$. We take this change to
be of a Hebbian nature in that it is proportional to the intensity of the
pre-synaptic input $S_i^\mu $ and to the desired output 
$\sigma^\mu_b $. This so-called Hebbian term, $\Delta J_i\propto
\sigma^\mu_b S_i^\mu $, could in principle be modulated by a series of other
processes, increasing or decreasing its importance, in order to enable the
system to learn more efficiently. This modulation can be represented by the
introduction of a modulation function $F$, such that now $\Delta J_i\propto
F\sigma^\mu_b S_i^\mu$ takes into account other factors of which at this
point we have no information.

While {\it ad hoc } algorithm building, i.e. choosing $F$, calls for
intuition, previous experience and some luck, the constructive nature of the
optimization procedure furnishes a set of variables ${\cal Z}$, on which the
modulation function depends as well as the function $F$ itself. If
optimization is carried under no restrictions, the set ${\cal Z}$ will
include all the variables which, if known, would contribute to achieve
optimal generalization. Several of them, call it the set ${\cal H}$, will
not be accessible or `hidden', while the remaining variables, belonging to ${\cal V}$,
are  accessible or `visible'. That is: ${\cal Z=H\cup V}$. The set ${\cal V}$ is the pool of
available information referred to in the previous section. Availability
conditions may, however, limit the set ${\cal V}$, thereby restricting the
learning scenario, leading to suboptimal learning conditions.

We mention what seems to be just a silly technical point, but will be seen
to be relevant in section (3). It concerns the presence of the correlation
term $\sigma^\mu_b S_i^\mu $ in the change of $J_i$. One could very well
optimize learning algorithms within the class of changes $\Delta
J_i=WS_i^\mu $, modulated by an {\it a priori} unknown function $W$,
obtaining exactly the same results as before. The optimal algorithms work 
neither by pure correlation nor error correction, they might resemble
both types of paradigms in different proportions in different stages of the
learning process \footnote{For the perceptron, at the early 
stages the algorithm is simple Hebb or pure correlation, while it 
resembles an error correcting (relaxation) algorithm later on. }.

\subsection{The Optimal Modulation Function.}

The evolution we want to discuss is restricted to the possible changes in
the function $F$ and the set ${\cal V}$, that is to the possible modulation
mechanisms and to their overall impact on the generalization ability of
systems restricted to Hebbian synaptic modification. The general form of the
optimal modulation function is then 
$$
F_{{\cal H|V}}^{opt}=J<\frac b\rho -h>_{{\cal H|V}} 
$$
proportional to the expectation value of the difference between the fields
averaged over the unknown quantities (${\cal H}$), given the pool of
available data (${\cal V}$). This form holds for the
linear, boolean and reverse wedge perceptron as well as related
architectures such as the tree parity and committee machines. For fully
connected architecture, although optimization has not been totally
completed, preliminary results show that the main feature still holds, that
is, the modulation functions are still expected values of the form $<...>_{
{\cal H|V}}$

Learning in the presence of restricted information can be studied by
accordingly limiting the set of available variables ${\cal V}$. By starting
with an empty ${\cal V}$ and sequentially promoting different members from $%
{\cal H}$ to ${\cal V}$ several different trajectories in the space of
algorithms can be defined. All lead from the simple Hebbian ($F=1$) 
to the fully
modulated optimal algorithm. An increase in the algorithms' complexity
occurs along each trajectory. We can rule out a trajectory to be a product
of `natural evolution' if there is a single step where the performance fails
to improve, on the grounds of a cost-benefit argument.

\subsection{Common Features of Optimal Modulation Functions}

Optimal algorithms, independently of the machine architectures so far
studied, share some characteristics.

First of all, the synaptic change depends on the example through the field $%
h $; not only through its sign $\sigma _h${\ \footnote{%
in the case of multilayer (ML) networks, this should read: fields $\{h_i\}$
and total output sign $\sigma(\{h_i\})$}, which determines if the student
agrees or not with the teacher on that particular instance; but also through
its absolute value, }$|h|$. This is used in evaluating the importance of any
`mismatch' between the answer expected by the student and the actual
teacher's version. A `small' $|h|$ may reduce the importance of an error,
whereas a `large' $|h|$ could indicate a particularly important example,
with a potentially high value of information. The scale in which `small' is
distinguished from `large' however 
is not the same throughout the learning process 
but can depend on several factors.

The modulation function is time dependent, meaning that the 
optimal annealing is built into the
modulation function. This is better described in terms of
performance dependence (Kinouchi and Caticha 1993)
rather than time. While the learning of a stationary
rule takes place, the generalization error decreases monotonically, and
therefore time and performance are interchangeable. However, if the learner
has to adapt to a time dependent rule then
time duration of the learning process looses importance. Performance though,
if and when it can be estimated at all, will still remain of value in
determining the amount of effective learning the machine has undergone.
It is fundamental to note that the main role of the performance determination, 
and therefore of the overlap $\rho$,
is in establishing the relevant scale of the field $h$, and thereby aid in
gauging the importance of the `mismatch 'or surprise in having expected a
different answer from that of the teacher.

Learning in the presence of noise 
(Biehl {\it et al} 1995, Copelli {\it et al} 1996, Heskes 1994)
introduces a host of interesting
variations. We will restrict to the case of output or multiplicative noise
which is characterized by a single parameter $\chi $, the (independent)
probability of a training label being inverted. Under this conditions a new
feature sets in and can be dubbed `confidence'. The importance of a surprise
may be watered down if the teacher is not reliable and there is trade off
between surprise and confidence. The cross over from being surprised to not
trusting the teacher depends both on the estimated performance and the noise
level \footnote{
`Confidence' also appears in the ML case, as different branches can, under
optimal conditions, determine which branch ought to be mostly blamed for an
overall error or equivalently having the least confidence in its partial
answer, even in the absence of noise.} $\chi $.

Of the different features on which the modulation function depends, we now
ask: in what order will they appear in a successive construction of the set $%
{\cal V}$? The set ${\cal Z}$ of relevant variables is determined by
unrestricted optimization. For the boolean perceptron we have $\{b,
\sigma_b, \xi, h, \sigma _h, \rho, \chi \}$. 
The teacher internal field $b$
is certainly not available, although in the conditions of supervised
learning, its sign $\sigma _b$, or at least a noise corrupted version of it,
 $%
\xi $ is. The values of the overall internal post-synaptic field $h $ or the
student network output $\sigma _h $ could be used. The average performance or
generalization error $e_g $ or the overlap $\rho $ or its estimate
may be present.
The noise level $\chi \ $, just an estimate of it (Biehl {\it et al} 1995)
or at least the knowledge of the existence of noise could also
be available \footnote{%
In a more general setting, the distribution of examples should be taken into
account, since biases in the inputs will certainly interfere with what
student network is considered optimal. Here we will restrict to the case of
uniformly, independent, randomly distributed examples.}.

The suboptimization in the presence of a different set of variables may lead
to differential performances. But not always. As one of the most interesting
cases, we consider the slightest increase in complexity that a pure
unmodulated Hebbian learning algorithm could undergo. The putative improved
algorithms can be obtained by finding the optimal modulation functions among
the possible $F(h,\sigma _b)$\footnote{In this case it depends on the
stability $\Delta =h\sigma _b\ $or $\Delta = h\xi \ $in the presence of noise}, 
$F(\rho ,\sigma _b)$ or still $F(\chi ,\sigma _b)$. It
turns out that modulating with $F(\Delta )\ $ leads to an improvement over
plain Hebbian 
\footnote{That is easy, 
since several $h$ dependent algorithms can be devised that
are nonoptimal improvements over the pure unmodulated
Hebbian algorithm. }.
But the best possible modulation within
the class of functions that depend only on $e_g\ $or $\chi \ $is $F=1$ (!) 
( Copelli {\it et al} 1996),
i.e. the information they bring is, at this point in the algorithm development
process, irrelevant. There is no advantage in developing hardware to measure
any of those if the necessary hardware to measure $\Delta \ $ is not already
there. The inclusion of the performance $e_{g\ }$leads to an improvement
once $\Delta \ $is available. This translates into an annealed algorithm. At this point, inclusion of the noise level
will lead into further improvement of the learning algorithm. The confidence, or lack of it, in the
supervised information leads to a possible rejection of outliers. Therefore
the performance  ordering
${\cal Q}(\xi ) = {\cal Q}(\xi, \rho) < {\cal Q}(\xi, h)<{\cal Q}(\xi, h, \rho)$
suggests a time ordering for the construction of ${\cal V}$ to be $(\xi) \rightarrow (\xi, h) \rightarrow (\xi, h, \rho)$, and not $(\xi) \rightarrow (\xi, \rho) \rightarrow (\xi, h, \rho)$.

The next section will try to make a parallel between these ideas and some
biological examples. Before that, we point out the physical meaning of these
variables. While $\Delta (<0)>0\ $signals (dis)agreement of the student with
the supervisor, $|\Delta |\ $indicates how sure was the student in
predicting the answer. A large negative $\Delta \ $indicates a big,
surprising mistake. The capacity of measuring it is tantamount to being able
to be surprised, whereas the possibility of modulating learning with a $%
\Delta \ $depending function implies in the use of the expected vs. actual
answer mismatch in learning. The influence of performance on the learning
algorithm is done by furnishing the appropriate scale in which surprises are
to be judged. A big surprise of an unexperienced learner should not be as
important as that of a more advanced student. Neither a big surprise will
have the same effect in a gullible student as in one which knows that the
incoming information is not completely reliable.

Practical algorithm builders may very well complain that these variables,
such as $\rho \ $or $\chi \ $are not available to the student network at all
but only certain estimations can at best be made. And in doing so they will
discover the main point we want to stress. Optimization methods will
determine firstly, the limits of learning and secondly, the possible
reliance on some usually hidden quantities in order to attain them. This
reliance, rather than hampering the utility of the optimization ideas, is
one of its main results. By showing which these variables are, it provides a
necessary pressure to develop algorithms or hardware which permit their more
efficient determination
\footnote{The creation of estimators for the hidden parameters
can be viewed as a second order learning process: learning
to learn more efficiently (Kinouchi and Caticha 1993, Biehl {\it et al} 1995,
Coehn and Jain 1994)}
This is the most obvious consequence. 
Whether similar ideas apply to  
biological evolution is an intriguing thought. Hints that this may be so
will be considered in the next section.

\section{Some examples drawn from Biology}

\subsection{Disclaimer}

Life in the real world is not as clear cut as in the simple laboratories of
the theories reviewed above. In trying to interpret experimental data, a
translation of the mathematical language we have employed is necessary. This
translation can be done at different hierarchical levels, ranging from the cellular
to the behavioral levels.

There is an extensive literature regarding receptive field and synaptic
plasticity, (e.g. Cruikshank and Wienberger 1996, Churchland and Sejnowski 1996)
and references therein. We
certainly can make no attempt at any comprehensive review, but will recall
some examples that seem to be natural in the light of previous sections.
By selectively choosing only some of those that seem to be on our side we do
not wish to imply lack of different mechanisms nor exclude other
laws. The critical interpretation of experimental work is left to the
experts, and we only use their conclusions. We also claim no final
conclusion, only present a set of experimentally backed up hints about the
relevance of the previous theoretical results.

\subsection{Examples}

First of all there is the question about the Hebbian nature of synaptic
plasticity or of its extensions, such as the Stent-Hebb law
e.g. (Churchland and Sejnowski 1996).The sheer variety of plastic neural
circuitry suggests that there very well may be other than just Hebbian-like mechanisms  to
encode information in the intercellular interactions.
Standard Hebbian theories hold that it is the pre-post-synaptical neural activity
correlation of activities that governs the change of synaptic
efficiencies. The necessity as well as sufficiency of Hebbian mechanisms to explain  a set of experiments has been satisfactoraly established. There is, however, in some other cases evidence for neuromodulated synaptic plasticity which transcends mere correlation mechanisms . 

The first example we deal with is the experiment of (Carew, Hawkins, Abrams and Kandel, 1984)
as discussed by (Cruikshank and Wienberger 1996).
In this experiment in the {\it Aplysia},
(a) the increase of the synaptic coupling between a pre-synaptic sensory
neuron and a post-synaptic motor neuron was induced by correlated activity
in both neurons, suggesting a Hebbian mechanism. This was obtained under
sensorial stimulation, which caused strong activity of the motor neuron.

To verify sufficiency they tested (b) the effect of simple activity
correlation. This was obtained by inducing post-synaptic depolarization, and
therefore activity, through the injection of current into the cell, in the
absence of external stimulation. Correlation was thus shown insufficient,
since alone it did not induce synaptic strengthening. Finally, in experiment
(c) its necessity was also discarded by showing an increase in synaptic
efficiency in the conditions of the first experiment modified by
hyperpolarizing the post-synaptic neuron and thus eliminating the activity
correlation.

Without the modulation due to behavioral context no significant plasticity 
occurs in experiment (b). This would be represented in our `modulated Hebbian' model by a small or absent modulation function $F$, which could be due to a behavioral feedback. We want to stress 
the difference between what has been called Hebbian in Physics and Physiology. 
Physicists have used the desired response $\sigma_b$ instead of the actual postsynaptic activity $\sigma_J$ in defining the Hebbian change. This difference is irrelevant for clamped neuron experiments. Synaptic changes, such as the ones described in section 2, proportional to $\delta S_i \sigma_J$, where $\delta=\sigma_J \sigma_b$  have been termed {\it predictive Hebbian\/} learning (Montague, Dayan, Person and Sejnowski, 1995).

Another set of experiments (Ahissar {\it et al.} 1992) deals with the role of 
behavioral context in the changes of the functional connection ({\it fc}) 
between neurons in the auditory cortex of monkeys. Instead of measuring 
single synaptic plasticity, they obtained the {\it fc}, which is the 
effective interaction that arises from contributions from all 
possible pathways between the two neurons, from the 
Cross Correlation Histograms (CCH) technique. The ratio of the {\it fc} 
after and before the conditioning, 
which they termed the Strengthening Factor (SF), was plotted against
the Contingency Factor (CF), the integrated coactivity (induced gain) during 
conditioning, obtained from the CCHs, divided by its value before 
conditioning. Although both in the presence of behavioral
relevant context and in its absence, SF and CF were positively correlated,
in the former case this relation was much more striking. Their conclusion, that in a behavioral relevant context, coactivity leads to a much higher {\it fc} change than when behaviorally irrelevant, is in accordance to what we discussed in section (2).

Among other possible examples, we mention, very briefly, that the influence of surprise in synaptic plasticity in the cerebellum of monkeys, has been reported by Gilbert and Thach (Kandel and Schwartz 1986).

We conclude by mentioning a few facts, very suggestive at the light of what 
we expounded in the previous sections, that occur, not at the cellular level,
but at neuropsychological one.
Several works (e.g. Kandel and Schwartz 1986, Shallice 1988)
have pointed the amygdala as responsible for identifying the `mismatch' or surprise element in the processing of new information.
Work on prefrontal syndrome patients has, on the other hand, indicated the role of the prefrontal lobe in evaluating performance levels related to working memory (`online') procedures (e.g. Williams and Golman-Rakic 1996)
have been reported to stick to strategies that were once successful ({\it perseverance} effect)
even if its clear, to a normal control, that the underlying rule has 
changed. Perseverance can be attributed to the lack of feedback from the working memory self-evaluation mechanism.

By excluding different variables in the set ${\cal V}$,
different types of lesions can be modeled in a feedforward net. 
The degradation of the performance ${\cal Q}$ when the 
self-evaluation module is ineffective shows `perseverance' 
effects when learning rules that may change unexpectedly in time. 
This is in contrast to the adaptive optimal algorithm which detects 
poor performance and can effectively start relearning once the rule changes.
We stress the fact that the optimal perceptron has not been built 
explicitly to present these perseverance effects when lesioned, but that 
it is uniquely deduced from the probability distribution associated 
to the task at hand by the sole requirement of having maximal 
generalization ability (Kinouchi and Caticha 1993, 1993b)
for its particular architecture (see also (Changeux 1992, Levine, Leven and Prueitt 1992)).

On the basis of our earlier remarks it is clear that the surprise measuring hardware is expected to be an {\it older}
structure than the performance measuring hardware. That this `time ordering' between the amygdala and the prefrontal lobe holds, is well known and therefore may come
as no surprise ( Kandel and Schwartz 1986).
That this is in accord with the theory developed for such simple
systems as the feedforward networks here discussed is where we think the real surprise lies. Whether other time ordering sequences can be thus identified will be the subject of future studies. 

Finally at an even more macroscopic level, we mention the 
Rescorla-Wagner model (Rescorla and Wagner 1972, Gluck and Bower 1988, Gluck 1991), 
widely studied in animal
psychology and also used to model human 
categorization, to support the study of these networks. Mathematically,  
the Rescorla-Wagner model is a simple perceptron, since decisions are 
based on weighted sums of signals, learning with an Adaline algorithm. 
The reason this is sensible is that, although at the 
microscopic level we have to deal with extremely complex 
networks, humans frequently use simple algorithms 
for inference and detection of statistical regularities. 
There is evidence that humans fail on some nonlinearly separable tasks (Thorpe, O'Reagan and Pouget 1990)
of an automatic and non-linguistic nature, suggesting that modeling 
problem-solving mechanisms by even a simple perceptron 
may be relevant in these cases. Since the optimal algorithm perceptron takes into
account the amount of surprise, stage of learning and a confidence measure
in order to learn effectively, we suggest that modeling
at macroscopic levels by perceptron architectures can only be ruled out 
after the inclusion of these properties into the model, otherwise, failure to
reproduce experimental results may lie somewhere else and not in the simple
network.
This work was supported by CNPq and Capes.

\small
{\bf REFERENCES}
\begin{itemize}
\item[]
Ahissar E., Vaadia E., Ahissar M, Bergman H., Arieli A. and Abeles M.,
{\em Science\/} {\bf 257} 1412 (1992).

\item[] 
Biehl M., Riegler P. and Stechert M.,
{\em Phys. Rev. E\/} {\bf 52} R4624 (1995).

\item[]
Biehl M. and Schwarze H., 
{\em J. Phys. A: Math. Gen.\/} {\bf 26} 2651 (1993).

\item[]
Carew T., Hawkins R., Abrams T., and Kandel E.
{\em J. Neurosci. \/} {\bf 4} 1217 (1984)

\item[]
Changeux J. P. 
{\em La Recherche} 244 {\bf 23} 705 (1992)

\item[]
Chen D. S. and Jain R. C.,
{\em IEEE Trans. Neural Networks\/} {\bf 5} 467 (1994).

\item[]
Copelli M. and Caticha N.,
{\em J. Phys. A: Math. Gen.\/} {\bf 28} 1615 (1995).

\item[]
Copelli M., Kinouchi O. and Caticha N.,
 {\em Phys. Rev. E\/} {\bf 53} 6341 (1996).

\item[]
Copelli M., Eichhorn R., Kinouchi O., Biehl M.
Simonetti R., Riegler P. and Caticha N.,
{\em Europhys. Lett}. {\bf 37 (6)}, pp. 427 (1997)

\item[]
Churchland P. and Sejnowski
{\em The Computational Brain} (MIT Books; Cambridge Mass) (1996)

\item[]
Cruikshank S. and Weinberger N.
{\em Brain Research Review} {\bf 22} 191 (1996)

\item[]
Gluck M. A. and Bower G. H.,
{\em J. Exp. Psychology: General\/} {\bf 117} 227 (1988).

\item[]
Gluck M. A.,
{\em Psychological Science\/} {\bf 2} 50 (1991).

\item[]
Heskes T., 
in {\em Proceedings of the ZiF Conference on
Adaptive Behavior and Learning\/}, Eds. J Dean ,
H. Cruse and H. Ritter (University of Bielefeld, Bielefeld,
Germany, 1994).


\item[]
Kandel E. R. and Schwartz J. H. (Eds.),
{\em Principles of Neural Science\/} (Elsevier, Holland, 1986).

\item[]
Kinouchi O. and Caticha N.,
{\em J. Phys. A: Math. Gen.\/} {\bf 25} 6243 (1992).

\item[]
Kinouchi O. and Caticha N.,
{\em J. Phys. A: Math. Gen.\/} {\bf 26} 6161 (1993).

\item[]
Kinouchi O. and Caticha N.,
in {\em Proceedings of the World Congress on Neural Networks -
WCNN'93} pg. 381-384 vol. III (Portland - 1993).

\item[]
Kinouchi O. and Caticha N.,
{\em Physical Review E\/} {\bf 54} R54 (1996).

\item[]
Levine D.S., Leven S. and Prueitt P.S. 
 in {\em Motivation, Emotion and Goal direction in Neural Nets}, Levine D.S. and Leven S. Eds. ( Hillsdale, NJ; Erlbaum) (1992)

\item[]
Montague P R, Dayan P, Person C. and Sejnowski {\em Nature \/} {\bf 377} 725 (1995)

\item[]
Opper M. 
{\em Phys. Rev. Lett.  \/} {\bf 77} 4671  (1996)

\item[]
Opper M. and Haussler D.,
in {\em Proceedings of the IVth Annual Workshop
on Computational Learning Theory (COLT91)\/},
(Morgan Kaufmann, San Mateo, 1991).

\item[]
Opper M. and Kinzel W.,
in {\em Physics of Neural Networks III\/} , Eds. E. Domany,
J. L.Van Hemmen and K. Schulten (Springer, Berlin, 1994).

\item[]
Opper M. and Winter O. 
{\em Phys. Rev. Lett.  \/} {\bf 76} 1964  (1996)

\item[]
Rescorla R. A. and Wagner A. R.,
in {\em Classical conditioning: II. Current research 
and theory\/}, Eds. A. H. Black and W. F. Prokasy
(Appleton-Century-Crofts, New York, 1972).

\item[]
Shallice T. 
{\em From Neuropsychology to Mental Structure}
(Cambridge; Cambridge Univ. Press) (1988)

\item[]
Simonetti R. and Caticha N.,
{\em J. Phys. A: Math. Gen.\/} {\bf 29} 4859 (1996).

\item[]
Sutton R. S. and Barto A. G.,
{\em Psychological Review\/} {\bf 88} 135 (1981).

\item[]
Thorpe S. J., O'Reagan K. and Pouget A.,
in {\em Neural Netwoks: From Models to Applications\/},
Eds. L. Personnaz and G. Dreyfus
(IDSET, Paris, 1990).

\item[]
Van den Broeck C. and Reimann P., 
{\em Phys. Rev. Lett.\/} {\bf 76} 2188 (1996).

\item[]
Watkin T. L. H.,
{\em Europhys. Lett.\/} {\bf 21} 871 (1993).

\item[]
Watkin T. L. H., Rau A. and Biehl M.,
{\em Rev. Mod. Phys.\/} {\bf 65} 499 (1993).

\item[]
Williams, G. V. and Goldman-Rakic P. S. 
{\em Nature} {\bf 376} 572 (1996)

\item[]
Winter O., Lautrup B. and Zang J-B.,
{\em Phys. Rev. E \/} {\bf 55} 936  (1997).

\end{itemize}
\end{document}